\documentclass[aps,prb,amsfonts,amssymb,twocolumn,amsmath,floatfix,showpacs]{revtex4-1}
\usepackage{amsbsy}
\usepackage{subfigure}
\usepackage[dvips]{color}
\usepackage[dvips]{graphics}
\usepackage{graphicx} 
\usepackage{bm} 
\usepackage[T2A]{fontenc}
\usepackage[cp1251]{inputenc}
\usepackage[english]{babel}

\newcommand{\be}{\begin{equation}}
\newcommand{\ee}{\end{equation}}
\newcommand{\bel}{\begin{align}}
\newcommand{\eel}{\end{align}}
\newcommand{\bem}{\begin{multline}}
\newcommand{\eem}{\end{multline}}
\newcommand{\beq}{\begin{equation}}
\newcommand{\eeq}{\end{equation}}
\newcommand{\bea}{\begin{eqnarray}}
\newcommand{\eea}{\end{eqnarray}}

\begin{document}
\title{Proximity-reduced range of internal phase differences in double Josephson junctions with closely spaced interfaces}

\author{Yu.\,S.~Barash}

\affiliation{Institute of Solid State Physics of the Russian Academy of Sciences, 
Chernogolovka, Moscow District, 2 Academician Ossipyan str., 142432 Russia}

\date{January 21, 2018}

\begin{abstract}
A substantial influence of the proximity and pair breaking effects on the range of internal phase differences is shown to
take place in symmetric double Josephson junctions with closely spaced interfaces and to affect the evolution of 
the supercurrent $j$ with the changing central lead's~length $L$. If the phase difference $\phi$ between~the external leads 
is controlled and $L$ exceeds~a~few~coherence lengths, the regime~of interchanging modes is established. The 
range of the~phase differences across the two individual interfaces is reduced with decreasing $L$, and the states of 
the higher energy mode are gradually eliminated. With a further decrease of $L$ the regime of interchanging modes is destroyed 
along with the asymmetric mode. The conventional single junction current-phase relation $j(\phi)$ is eventually established and 
the condensate states' doubling is fully removed at very small $L$.
\end{abstract}


\maketitle

\section{Introduction} 

Static and dynamic couplings of two closely spaced junctions can play an important role in mesoscopic 
systems of superconducting electronics \cite{Kao1977_2,HansenLindelof1984,Likharev1984,Blackburn1987,Smith1990,Blamire1994,%
Kupriyanov1999,Goldobin1999,Nevirkovets1999,Golubov2000,Brinkman2001,Ishikawa2001,Blamire2006,Luczka2012,Linder2017}. Two 
Josephson junctions connected in series are particularly linked to each other by the equality of the flowing currents. If 
distance $L$ between the junctions significantly exceeds the coherence length $\xi(T)$, then, in the absence of the magnetic 
effects, the junctions' coupling is negligible. However, in the opposite case $L\ll\xi(T)$, the proximity effects can strongly 
influence the transport processes, including the dc Josephson current.

A double Josephson junction with two thin interfaces is characterized by the phase differences $\chi_{1,2}$ across~them. 
The phase difference $\phi$ between the external leads generally reveals less information.~At~fixed $\phi$ the dc Josephson current 
still remains uncertain to some extent. For example, let the phase incursion over the central lead be negligible with the relation 
$\phi=\chi_1+\chi_2$ holding. Taking $\chi_1=\chi_2+2\pi n$ with integer $n$ for symmetric double junctions, one gets $\chi_1=
\frac{\phi}2+\pi n$ and transforms the single junction $2\pi$-periodic current-phase relation $j(\chi_1)$ into two different 
$4\pi$-periodic modes $j(\frac{\phi}2)$ and $j(\frac{\phi}2+\pi)$, with respect to $\phi$. Either mode describes, in particular,
the supercurrent sign change, when $\phi\to \phi+2\pi$ due to the coordinated variations of $\chi_{1,2}$ by $\pi$. If only 
one of the $\chi_{1,2}$ varies by $2\pi$ and induces the change $\phi\to \phi+2\pi$, one should simultaneously switch over to 
another mode to keep the current unchanged. Therefore, the current is at least a double-valued function of $\phi$, if 
$\chi_{1,2}$ are controlled in experiments independently as can occur at a sufficiently~large~$L$.

An alternative experimental possibility is to control $\phi$ allowing $\chi_{1,2}$ to take on the most preferable equilibrium values. 
The energetically favorable mode is formed by $j(\frac{\phi}2)$ within the periods $(4n-1)\pi\le\phi\le(4n+1)\pi$, and 
$j(\frac{\phi}2+\pi)$ at $(4n+1)\pi\le\phi\le(4n+3)\pi$. Here, unlike the junctions containing Majorana fermions 
\cite{Kitaev2001,Kane2009,Beenakker2013,Alicea2012,Molenkamp2016,Molenkamp2017}, the two originally  $4\pi$-periodic states with 
different~currents~$j(\frac{\phi}2)$~and~$j(\frac{\phi}2+\pi)$ get interchanged, when the phase $\phi$ is advanced by $2\pi$. 
Omitting here possible ``undercooling'' and ``overheating'' of the states at the transition, one gets a regime 
of interchanging modes described by a $2\pi$-periodic sawtooth-like current-phase relation with discontinuities at $\phi=(2n+1)\pi$\,
\cite{Luca2009}. The anharmonic relation, associated with the condensate states' doubling at given $\phi$, can be partially 
smoothed out by fluctuations, small junction asymmetries etc. \cite{Linder2017} 

At $L\ll\xi(T)$, the double Josephson junction, in fact, represents a single junction with a thin~interface~that includes the 
central region. Though only a sequential tunneling, rather than a direct one, is permitted, one could assume in this limit the 
regular single junction phase dependence~$j(\phi)$ on $\phi$. Although there is some experimental evidence supporting this issue
\cite{Blamire1994}, theoretical results diverge in respect~of it. The sawtooth current-phase relation has~been discussed at small 
$L$. \cite{Luca2009} The proximity effects, disregarded in \cite{Luca2009}, have been known to be important at $L\ll \xi$ and lead 
to a strongly phase-dependent order parameter in the central lead. \cite{Kupriyanov1988,Kupriyanov1999,Golubov2000,Ishikawa2001} 
Those microscopic studies resulted in the conventional single junction behavior at very small $L$, however without
taking the regime of interchanging modes into account. Finally, the results obtained within the Ginzburg-Landau (GL) approach, have 
shown no solutions at $L<\pi\xi(T)$ and, in particular, no single junction behavior. \cite{SolsZapata1996}

This paper develops a theory of symmetric double Josephson junctions within the GL approach with an improved interface description. 
An effective mutual impact of the internal phase differences, induced by interfacial proximity and pair breaking effects, will be 
identified and shown to result in their range being substantially reduced. The double Josephson junction with closely spaced 
interfaces is one of the simplest systems, where the effect occurs. As a consequence, a gradual destruction of the higher energy mode 
takes place with decreasing $L$. In particular, the state with $\chi_{1}=\chi_{2}=0$ will be discovered to occur at an arbitrary $L$, 
while the equilibrium state with $\chi_{1}=\chi_{2}=\pi$ - to exist only if  $L>L_{\pi}$. In the regime of interchanging modes, the 
abrupt change of the supercurrent in immediate vicinities of $\phi_n=(2n+1)\pi$ actually occurs continuously via the current-carrying
asymmetric states. Thus in the tunneling limit the symmetry $j(\pi-\chi)=j(\chi)$ allows one to associate the value~$\phi=\pi$~with 
$\chi_1$ and $\chi_2=\pi-\chi_1$ at all possible $\chi_1$, i.e., at any value $|j|\le j_c$. 
With a further decrease of $L$, the proximity is shown to reduce the order parameter in the central 
lead and the range of $\chi_{1,2}$ in such a way that it removes the regime of interchanging modes along with the asymmetric states, 
and eventually results in the single junction dependence $j(\phi)$ at all $\phi$.

\section{Description of the model}

\begin{figure}[t]
\centering
\includegraphics*[width=.8\columnwidth,clip=true]{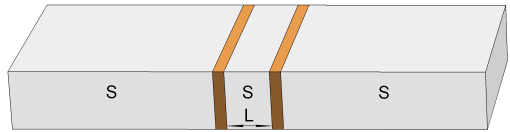}
\caption{Schematic diagram of the double junction} \label{fig: dj}
\end{figure}

Consider a symmetric double junction, which is made of the same superconducting material and contains two identical thin interfaces 
at a distance $L$, connected by the central superconducting lead (see Fig.\,\ref{fig: dj}). The interface thickness is on the order of or less than the 
zero-temperature coherence length $\xi_0$ considered to be zero within the GL theory. The length of the two external leads 
significantly exceeds the coherent length $\xi(T)$ and the magnetic penetration depth $\lambda(T)$. The one-dimensional spatial 
dependence of the order parameter is assumed, occurring, for example, when the transverse dimensions of all three electrodes are 
substantially less than $\xi(T)$ and $\lambda(T)$. The system's free energy is the~sum~of contributions from the interfaces and 
the bulk of the leads ${\cal F}=\sum{\cal F}_{p}+{\cal F}^{\text{int}}_{\frac{L}2}+{\cal F}^{\text{int}}_{-\frac{L}2}$. Here 
$p=1,2$ refer~to~the external electrodes, while $p=3$ refers to the central lead. One gets per unit area of the cross section
\be
\!\!{\cal F}_p\!=\!\!\!\int\limits_{{\cal C}_p}\!\!dX\!
\left[\!K\left|\dfrac{d}{dX}\Psi(X)\right|^2\!\!\!+a\left|\Psi(X)\right|^2\!\!+\dfrac{b}{2}\left|\Psi(X)\right|^4\right].
\label{Fb1}
\ee
For the interfaces placed at $X=\pm L/2$, the integration periods ${\cal C}_p$ for $p=1,2,3$ should be taken as 
$(-\infty,-L/2)$, $(L/2,\infty)$ and $(-L/2,L/2)$, respectively.

The interfacial free energy per unit area is
\begin{equation}
{\cal F}^{\text{int}}_{\pm\frac{L}2}=g_{J}\left|\Psi_{\pm\frac{L}{2}+}-\Psi_{\pm\frac{L}{2}-}\right|^2\!\!
+g\left(\left|\Psi_{\pm\frac{L}{2}+}\right|^2\!\!+\left|\Psi_{\pm\frac{L}{2}-}\right|^2\right).
\label{fint1}
\end{equation}
The two invariants in \eqref{fint1} describe the Josephson coupling with the coupling constant $g_J$ and the
interfacial pair breaking $g>0$. For $0$-junctions considered below $g_J>0$.

The GL equation for the normalized absolute value of the order parameter $\Psi=(|a|/b)^{1/2}fe^{\mathtt{i}\varphi}$ 
takes the form
\begin{equation}
\dfrac{d^2f}{dx^2}-\dfrac{i^2}{f^3}+f-f^3=0.
\label{gleq2}
\end{equation}
Here $x=X/\xi(T)$, $\xi(T)=(K/|a|)^{1/2}$ and the dimensionless current density is $i=\frac{2}{3\sqrt{3}}
(j\big/j_{\text{dp}})$, where $j_{\text{dp}}=\bigl(8|e||a|^{3/2}K^{1/2}\bigr)\big/\bigl(3\sqrt{3}\hbar b\bigr)$ is the 
depairing current deep inside the superconducting leads. 

The boundary conditions for the complex order parameter, which follow from \eqref{Fb1} and \eqref{fint1}, agree with the 
microscopic results \cite{Golubov2004} near $T_c$, at all transparency values \cite{Barash2012,Barash2012_2,Barash2014_3}. 
Introducing $l=L/\xi(T)$, one gets at $x=l/2$:
\begin{align}
&\left(\dfrac{df}{dx}\right)_{l/2\pm0}\!\!=
\pm\Bigl(g_\delta+g_\ell\Bigr)f_{l/2\pm0}\mp g_\ell\cos\chi f_{l/2\mp0},
\label{bc1} \\
&i=-\,f^2\left(\dfrac{d\varphi}{dx}+\dfrac{2\pi\xi(T)}{\Phi_0}A\right)
=g_{\ell}f_{l/2-0}f_{l/2+0}\sin\chi .
\label{joscurr1}
\end{align}
Here $\chi=\varphi\left(\frac{l}{2}-0\right)-\varphi\left(\frac{l}{2}+0\right)$, $\Phi_0=\frac{\pi\hbar c}{|e|}$
and the dimensionless coupling constants are $g_\ell=g_J\xi(T)/K$, $g_{\delta}=g\xi(T)/K$.

The boundary conditions \eqref{bc1} and the conservation of the supercurrent \eqref{joscurr1} allow the values $f_{l/2\pm0}$ 
on opposite sides of the interface between identical superconductors to differ from one another. In a single symmetric 
Josephson junction, $f(x)$ is usually continuous across the thin interface. However, the joint pair breaking by both end interfaces 
can more weaken the condensate density in the short central lead. The corresponding phase dependent jump $f_{l/2+0}-f_{l/2-0}>0$
allows superconductivity to survive in the central lead at $l\ll1$. The continuity of $f(x)$ across thin interfaces in double 
Josephson junctions is a distinctive feature of earlier theories that used the GL approach with the flawed boundary conditions 
for the order parameter \cite{Baratoff1975,Kao1977,SolsZapata1996}. Generally, those models are neither equivalent to the 
free energy \eqref{Fb1} and \eqref{fint1}, nor to the microscopic~results~near~$T_c$.~\cite{Kupriyanov1988,Kupriyanov1999,%
Golubov2000,Golubov2004}

There are a number of solutions that satisfy equation \eqref{gleq2}, the asymptotic conditions deep inside the external electrodes 
and the boundary conditions at $x=\pm l/2$ (see also Appendix A). The solutions with the preferred energies are assumed to have the 
extrema at $x=0,\,\pm l/2,\, \pm\infty$, or, when possible, only at $x=\pm l/2,\,\pm\infty$. The numerical simulations show that the 
symmetric solutions $f(x)=f(-x)$ with the internal phase differences $\chi_1=\chi_2+2\pi n=\chi$, occur in most cases considered 
below, except for close vicinities of $\phi_n=(2n+1)\pi$, where the asymmetric mode prevails, if~it~exists.

\section{Currentless states}

\begin{figure}[t]
\begin{center}
\begin{minipage}{.49\columnwidth}
\includegraphics*[width=.9\columnwidth,clip=true]{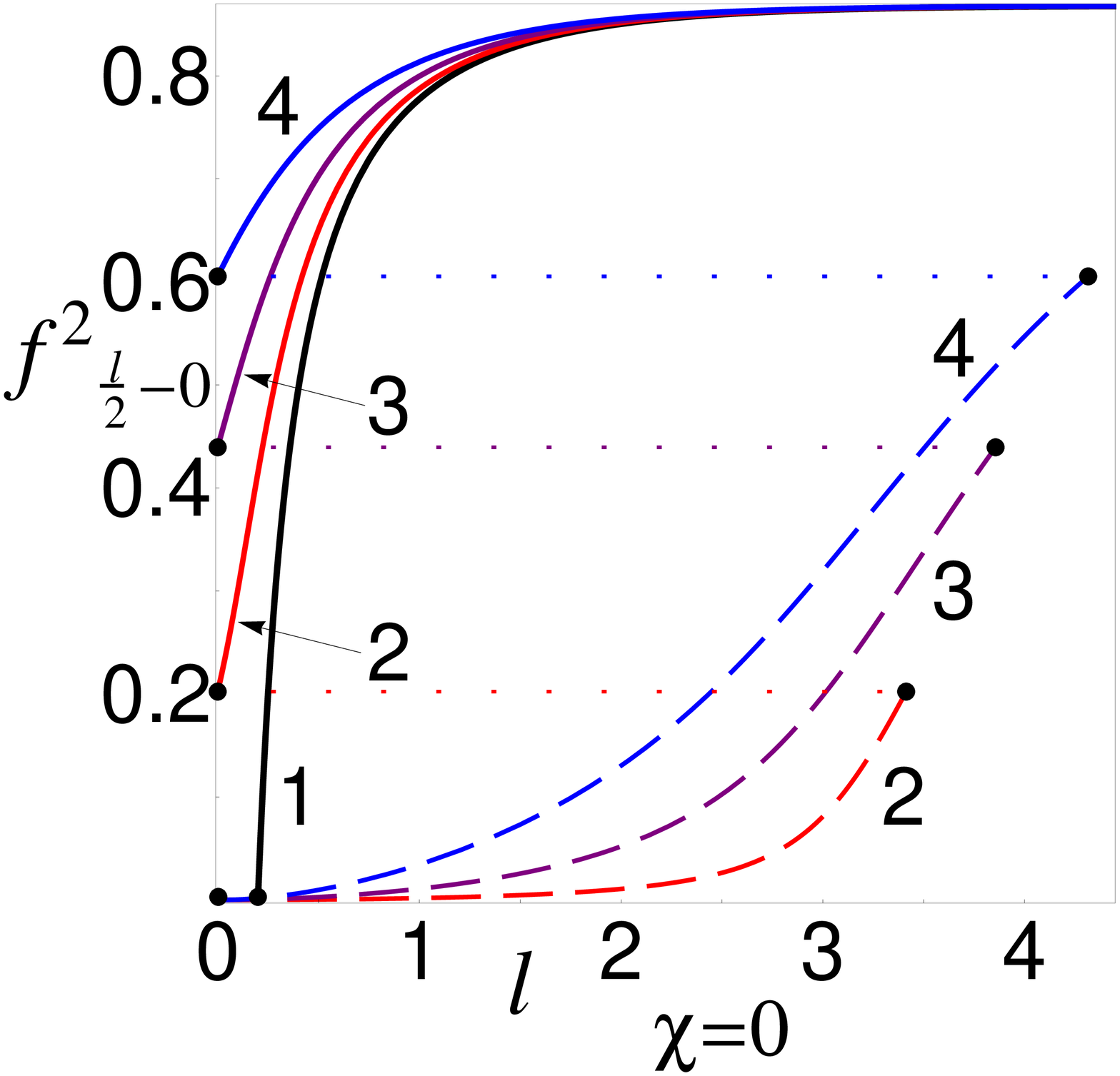}
\end{minipage}
\begin{minipage}{.49\columnwidth}
\includegraphics*[width=.97\columnwidth,clip=true]{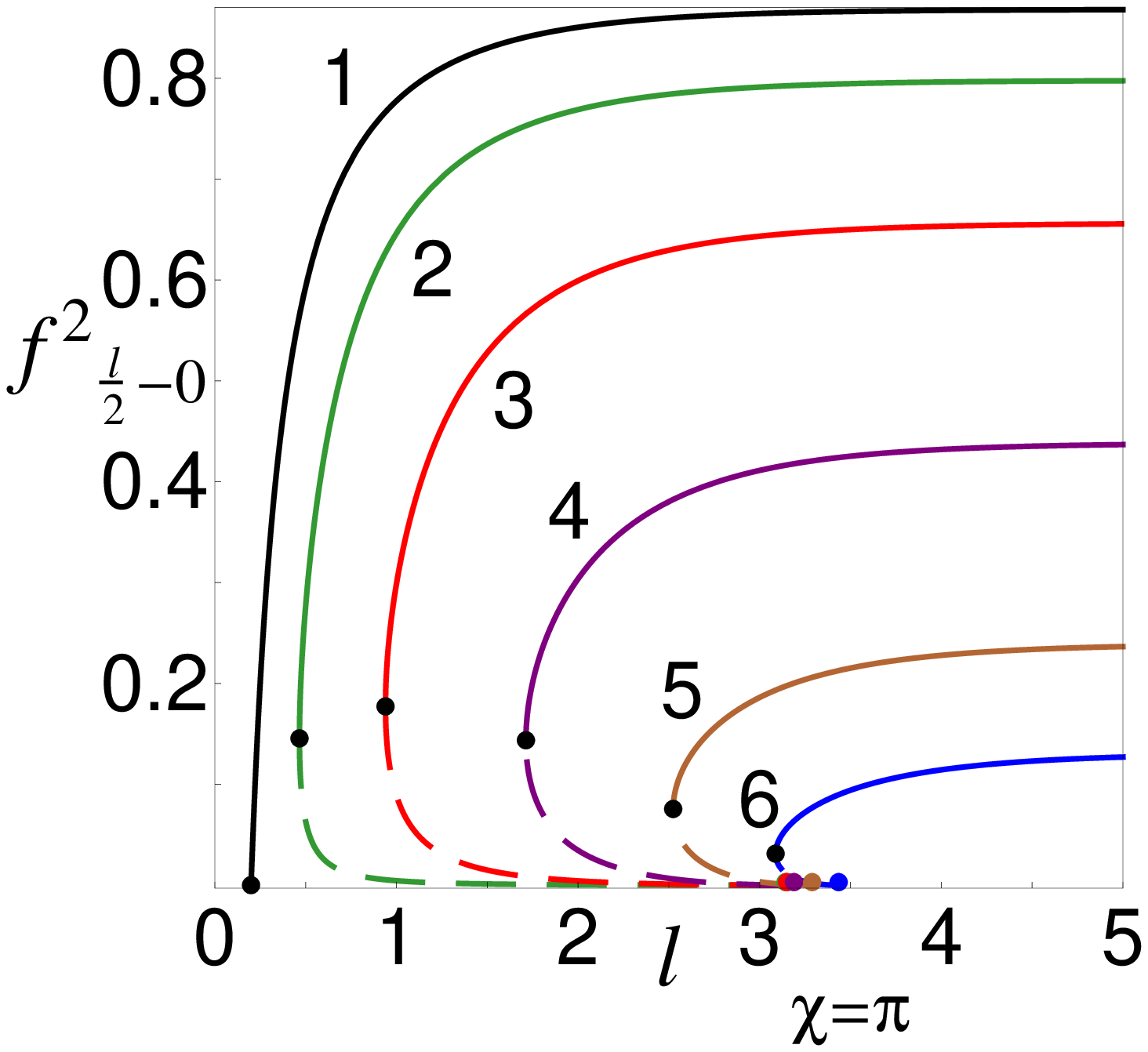}
\end{minipage}
\end{center}
\caption{$f^2_{l/2-0}$ as a function of $l$ 
at $\chi=0$ (left panel) and $\chi=\pi$ (right panel). 
Solid curves correspond to the energetically preferable states. 
{\it Left panel}: $\chi=0$,\, $g_\delta=0.1$ and  \,\,
(1)\, $g_{\ell}=0$\,\,
(2)\, $g_{\ell}=0.1$\,\,
(3)\, $g_{\ell}=0.3$,\,\, and \,\,
(4)\, $g_{\ell}=0.8$.
{\it Right panel}:
$\chi=\pi$,\, $g_\delta=0.1$ and \,\,
(1)\, $g_{\ell}=0$\,\,
(2)\, $g_{\ell}=0.03$\,\,
(3)\, $g_{\ell}=0.1$,\,\, 
(4)\, $g_{\ell}=0.25$,\,\, 
(5)\, $g_{\ell}=0.5$,\,\,and \,\,
(6)\, $g_\ell=0.8$.}
\label{fig:rminchi}
\end{figure}

The double junction's states with vanishing supercurrent at $\chi=\pi n$ allow the exact analytical description
(see Appendix \ref{sec: chi0pi} for details of the derivations). 
The quantity $f^2_{l/2-0}(\chi,g_\ell,g_\delta)$, taken at the boundary of the central lead, is 
depicted in Fig.\,\ref{fig:rminchi} as a function of $l$ at $\chi=0$ (the left panel) and 
$\chi=\pi$ (the right panel), for $g_\delta=0.1$ and various $g_\ell$. 
The solid curves describe energetically preferable solutions, while the dashed curves correspond to metastable states.

For an impenetrable wall ($g_\ell=0$) dependence on the phase difference vanishes, and the curves 1 in both panels in 
Fig.\,\ref{fig:rminchi} are identical. The free energy density of the sample placed between two impenetrable pair breaking 
walls is known to increase with decreasing $l$ due to the inverse proximity effects, and the transition to the normal metal 
state occurs at $L=2\xi(T)\arctan g_\delta$ \cite{ROZaitsev1965,Ginzburg1993,Barash2017}.~For $g_\delta=0.1$ one gets $l=0.199$. 
By contrast, for a nonzero Josephson coupling, the superconducting state with $\chi=0$ exists in the central electrode at any 
value of its length. The solid curves 2-4 in the left panel show that the quantity $f^2_{l/2-0}$ takes on its nonzero minimum 
value at $l=0$, unless $g_\ell\to0$ at $g_\delta\ne0$ (see also \eqref{calem142}).

At $\chi=0$, the two terms on the right-hand side of the boundary condition  $(df/dx)_{(l/2)-0}=-(g_\delta+g_\ell)
f_{(l/2)-0}+g_\ell f_{(l/2)+0}$ contain $f_{(l/2)-0}$ or $f_{(l/2)+0}$ as a factor and have opposite signs. 
If $f_{(l/2)-0}>g_\ell f_{(l/2)+0}/(g_\delta+g_\ell)$, the derivative is negative and $f(x)$ increases with decreasing $x$ up 
to $x=0$. Such solutions correspond to the solid curves 2 - 4 in the left panel of Fig.\,\ref{fig:rminchi}. If the equality  
$f_{(l/2)-0}=g_\ell f_{(l/2)+0}/(g_\delta+g_\ell)$ holds, the derivative $(df/dx)_{(l/2)-0}$ at the boundary of the central 
lead vanishes. There is also the solution of a different type, depicted by the dashed curves in the left panel of 
Fig.\,\ref{fig:rminchi}, for which $(df/dx)_{(l/2)-0}>0$ and $f(x)$ decreases, when $x$ goes down inside the central lead, and 
vanishes at $x=0$. Such a metastable solution, induced by the proximity to the external superconducting electrodes, has smaller 
values and satisfies the relation $f_{(l/2)-0}<g_\ell f_{(l/2)+0}/(g_\delta+g_\ell)$. 

Unlike the case $\chi=0$, the terms on the right-hand side of the boundary conditions \eqref{bc1} have identical sign at $\chi=\pi$.
Therefore, the condensate density decreases the nearer one gets to the interface irrespective of the relation between 
$f_{(l/2)-0}$ and $f_{(l/2)+0}$. As a result, for the state with $\chi=\pi$ to exist the length $l$ has to exceed the critical value 
$l_{\pi}(g_\ell,g_\delta)$. However, a disappearance of the equilibrium state with $\chi=\pi$ at $l<l_{\pi}(g_\ell,g_\delta)$ and 
$g_\ell\ne0$ is not accompanied by a transition to the normal metal state, in contrast to what takes place at $g_\ell\equiv0$.
 
The transition to the normal metal state of the system as a whole, with distant regions of the external electrodes, is
energetically unfavorable since the interfacial pair breaking is confined by the scale $\alt\xi(T)$. 
Were only the central electrode in the normal metal state, the boundary condition 
$(df/dx)_{(l/2)-0}=-(g_\delta+g_\ell)f_{(l/2)-0}-g_\ell f_{(l/2)+0}$ at $x=l/2-0$ and $\chi=\pi$ would result in $f_{(l/2)+0}=0$ 
once $g_\ell\ne0$. In this case one also gets $(df/dx)_{(l/2)+0}=0$ from the boundary condition on the opposite side of the 
interface.~These~two~equalities signify vanishing superconductivity throughout the external leads, which is not possible as stated 
above. Thus, $\chi=\pi$ is not the equilibrium value of $\chi$ under the conditions $l<l_{\pi}(g_\ell,g_\delta)$ and $g_\ell\ne0$, 
while superconductivity does exist due to the proximity to the external superconducting leads.

The metastable solutions at $\chi=\pi$, depicted by the dashed curves in the right panel of Fig.\,\ref{fig:rminchi},
are of the same type as the energetically preferable ones. They appear within the range 
$l_{\pi}(g_\ell,g_\delta)<l<l_{ps}(g_\ell,g_\delta)$.  At $l=l_{ps}(g_\ell,g_\delta)$ 
the metastable phase-slip centers arise on the central lead's end interfaces: $f_{\pm(l_{ps}/2-0)}=0$ (see Appendix 
\ref{sec: chi0pi}). 
In the tunneling limit $l_{ps}$ takes on its minimum value $l_{ps}(g_\ell\to0,g_\delta)=\pi$.
The points with coordinates $l=l_{ps}(g_\ell,g_\delta)$ and $f_{l/2-0}=0$ are marked in the right panel of 
Fig.\,\ref{fig:rminchi}.

The numerical study of the solutions shows that the left and right panels of Fig.\,\ref{fig:rminchi} 
represent the two main types of mapping of $f^2_{l/2-0}$. The transformation of one type into another 
with changing $\chi$ usually occurs some distance below $\chi=\pi/2$ within a noticeable interval~$\Delta\chi$.

\section{Double junctions with $l\ll1$}

Within~the~zeroth-order approximation in the small parameter $l\ll1$,~the~symmetric solution of the GL equation, complemented 
by the boundary conditions at the interfaces and the asymptotic conditions deep inside the external electrodes, satisfies, 
if $\cos\chi>0$, the relation $f_{l/2-0}=\dfrac{g_\ell\cos\chi}{g_\delta+g_\ell}f_{l/2+0}$, which 
leads to vanishing derivative $\left({df}/{dx}\right)_{l/2-0}$ (see Appendix \ref{sec: smalll}). However, the 
pair breaking effects do not allow the phase differences with $\cos\chi<0$ to be established in the equilibrium.

The condition $\cos\chi>0$ results in the allowed bands $-\pi/2+2\pi n<\chi<\pi/2+2\pi n$ and the forbidden gaps between them. 
Switching over to the $\phi$-dependence and disregarding the phase incursion over the central lead, one gets the same bands for 
the argument $\frac{\phi}2$ of the first mode, whereas the allowed and forbidden gaps are interchanged for the argument 
$\frac{\phi}2+\pi$ of the second mode. Combining the allowed bands of both modes, which are tightly adjoined to each other but do not 
overlap, results in the single-valued dependence on $\phi$ of the quantities in question, at all real $\phi$. Here the functions 
$\cos\chi$ and $\sin\chi$ should be replaced by $\cos\frac{\phi}2$ and $\sin\frac{\phi}2$, if $(4n-1)\pi\le\phi\le(4n+1)\pi$, and by 
$\cos(\frac{\phi}2+\pi)=-\cos\frac{\phi}2$ and $-\sin\frac{\phi}2$ in the case $(4n+1)\pi\le\phi\le(4n+3)\pi$.  Therefore, one 
obtains at any value of $\phi$
\begin{align}
&f_{l/2-0}=\frac{g_\ell|\cos\frac{\phi}2|}{g_\delta+g_\ell}f_{l/2+0}, \label{rel1}\\
i=g_\ell^{\text{eff}}&f^2_{l/2+0}\sin\phi, \quad g_\ell^{\text{eff}}=\dfrac{g_\ell^2}{2(g_\delta+g_\ell)},
\label{joscurr2}
\end{align}
where the right hand side in \eqref{joscurr1} has been used in \eqref{joscurr2}.
\begin{figure}[t]
\centering
\includegraphics*[width=.83\columnwidth,clip=true]{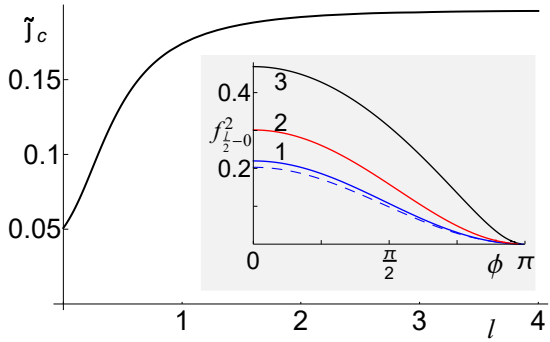}
\caption{Critical current as a function of $l$ at $g_{\ell}=0.1$, $g_{\delta}=0.1$.\,Inset: 
The quantity $t_{l/2-0}(\phi)$ at (1)\, $l=0.02$\,\, (2)\, $l=0.1$\,\, (3)\, $l=0.25$.}
\label{fig:cur1}
\end{figure}

Remarkably, the higher energy mode present to the full extent at large $l$ is completely destroyed in the limit of very small $l$ 
due to the proximity reduced range of the internal phase differences. While the~low~energy~mode~can be distorted at large $l$ by 
the ``undercooling'' and ``overheating'' processes, it is stabilized at small $l$. The total elimination of the condensate states' 
doubling at any given $\phi$ and the GL expression \eqref{joscurr2} for the supercurrent across the junction reduce
the double junction behavior in the limit $l\to0$ to that of a symmetric single junction with the effective Josephson 
coupling $g_\ell^{\text{eff}}$.

The supercurrent \eqref{joscurr2} decreases with $\phi$ at $\pi/2\le\phi\le\pi$ at the expense of the proximity-induced 
phase dependent factor $|\cos\frac{\phi}2|$ on the right hand side of \eqref{rel1}. Since in 
tunnel junctions $g_\ell\propto{\cal D}$, where $\cal D$ is the interface transmission coefficient, $g_\ell^{\text{eff}}\propto 
{\cal D}^2$, when $g_\delta\gg g_\ell$, and $g_\ell\propto {\cal D}$ in the opposite limit $g_\delta\ll g_\ell$, in agreement with 
the earlier microscopic results. \cite{Kupriyanov1988,Kupriyanov1999,Golubov2000} As follows from \eqref{joscurr2} and \eqref{bc1}, 
the effective interfacial pair breaking parameter, in the zeroth order in $l$, is $g_\delta^{\text{eff}}=
g_\delta(g_\delta+2g_{\ell})\big/(g_\delta+g_{\ell})$.

A strong suppression of the quantity $f_{l/2-0}$~in~a~close vicinity of $\phi=\pi$ and the supercurrent spatial uniformity 
entail a large gradient of the order-parameter phase. As the numerical study shows, this results, even at 
very small $l$, in a noticeable phase incursion over the central lead that violates the applicability of the zeroth-order 
approximation in $l$ near $\phi=\pi$. Although there are no discernible modifications near $\phi=\pi$ in 
\eqref{rel1} and \eqref{joscurr2}, the range of $\chi$ is more restricted so that only values at a distance below 
$\chi=\pi/2$ are permitted at~small~$l$.

The solid curves 1 - 3 in the inset in Fig.\,\ref{fig:cur1} show the numerical results for the phase-dependent order parameter 
squared $f^2_{l/2-0}$ taken at the central lead's end face. The dashed curve that corresponds to the right-hand side of \eqref{rel1} 
at $l=0.02$, coincides with~curve~1 with only a small percentage of deviation. ~Due~to~a~weak~dependence~on~$l$~of 
the order parameter $f_{l/2+0}$ on the opposite side of the interface, the dashed curves at $l=0.1$ and $l=0.25$ (not shown) 
almost coincide with the one presented for $l=0.02$ and, therefore, substantially deviate from the solid curves 2 and 3. Thus the 
relation \eqref{rel1}, justified at $l=0.02$ for the chosen set of parameters, is~violated~with increasing $l$ already~at $l=0.1$ 
and $l=0.25$. 
\begin{figure}[t]
\centering
\includegraphics*[width=.9\columnwidth,clip=true]{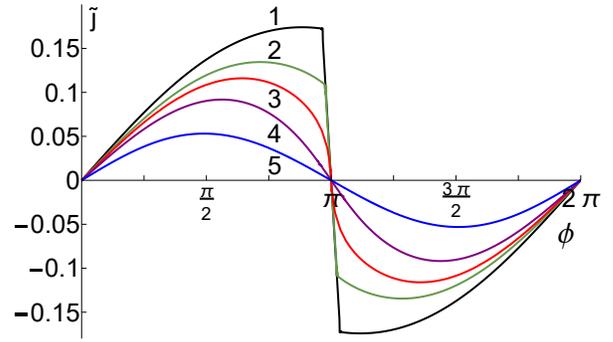}
\caption{Current-phase relations $\tilde\jmath(\phi)$ taken for $g_{\ell}=0.1$,\, $g_\delta=0.1$ and \,\, (1)\, $l=1$\,\,
(2)\, $l=0.5$\,\, (3)\, $l=0.38$\,\, (4)\, $l=0.25$\,\, (5)\, $l=0.02$.}
\label{fig:cur2}
\end{figure}

The current-phase relation $\tilde{\jmath}(\phi)$, taken at various $l$, is depicted in Fig.\,\ref{fig:cur2} for the  
supercurrent $\tilde{\jmath}=j/j_{\text{dp}}$ and the interfaces with $g_\ell=g_\delta=0.1$. The numerical results have been 
obtained by carrying out the evaluation of the supercurrent \eqref{joscurr1} with the consistent solutions of the model \eqref{Fb1}, 
\eqref{fint1}, including the phase incursion over the central lead.
For the given set of parameters, the asymmetric states, along with a noticeable abrupt change of the supercurrent in the 
vicinities of $\phi_n=(2n+1)\pi$, are fully destroyed by the pair breaking effects below $l\approx 0.36$. The curves 1 and 2 show 
that the regime of interchanging modes still takes place at $l=1$ and $l=0.5$. By contrast, the curve 5 for $l=0.02$ differs only by 
several percent from the one corresponding to the conventional single junction current-phase dependence \eqref{joscurr2}. 
Anharmonic contributions to $\tilde{\jmath}(\phi)$ are characteristic of the curves 3 and 4. The critical current of 
the double junction as a function of $l$ is shown in the main panel of Fig.\,\ref{fig:cur1}.

In conclusion, the double Josephson junctions with closely spaced interfaces have been theoretically \mbox{studied}.  With 
decreasing central lead's length $l$, the range of~the internal phase differences is shown to be gradually reduced.
At very small $l$, the condensate states' doubling at any given $\phi$ is fully removed and the single junction expression 
\eqref{joscurr2} describes the Josephson current.

\appendix

\section{Symmetric solutions of the GL equation}
\label{sec: sol}

For identifying the Josephson current 
(5), one should know the 
order parameter interface values $f_{(l/2)\pm0}$ as functions of the
phase difference $\chi$ and the length $l$, 
and of other GL theory's parameters. 
The simplest way to obtain the results is to make use of the first
integral of the GL equation (3). The 
quantity ${\cal E}$, defined as
\be
{\cal E}=\left(\dfrac{df(x)}{dx}\right)^2+\dfrac{i^2}{f^2(x)}+f^2(x)-\dfrac{1}{2}f^4(x),
\label{RPhi88gt}
\ee
is spatially constant inside each of the leads, when taken for the solutions of (3). 
Different values of $\cal E$ in different leads can appear due to the boundary conditions, which
follow from (1) and (2) and do not generally support the conservation of ${\cal E}$ through the 
interfaces.

Eq. \eqref{RPhi88gt} can be also rewritten in the form
\be
\left(\dfrac{df}{dx}\right)^2=\dfrac{1}{2f^2}(f^2-f_+^2)(f^2-f_d^2)(f^2-f_-^2).
\label{RPhi89gt}
\ee

The quantities $t_-=f_-^2$, $t_d=f_d^2$ and $t_+=f_+^2$ satisfy the
following set of equations
\begin{align}
t_-+t_d+t_+=2,&\qquad t_-t_dt_+=2i^2, \nonumber\\
t_dt_-+t_dt_++&t_-t_+=2{\cal E}.
\label{ttt1}
\end{align}

Solutions of equation \eqref{RPhi89gt} are characterized by three formal 
extrema $f_-,\,f_d,\,f_+$ with the vanishing first derivative $\frac{df}{dx}$.
In general, either all three roots $t_-,\, t_d$ and $t_+$ take on real values, or only one is 
real and two are the complex conjugate of each other. As the numerical study shows,
only real values are relevant for the given problem, and the case with three real minimums is also
excluded, at least for the set of parameters studied. As the left hand side of \eqref{RPhi89gt} 
takes on nonnegative values, there should be, therefore, one minimum (let it be $t_-$) and two 
maximums $t_+\ge t_d\ge t(x)$ among the three real roots. 

Symmetric analytical solutions of the GL equation (3) describe the order-parameter  
absolute value as a function of $l$ and $\chi\equiv\chi_1=\chi_2+2\pi n$, and satisfy the boundary conditions 
at $x=\pm l/2\pm0$ (see, e.g., (4)) as well as the asymptotic conditions deep inside the long external leads.
The energetically most favorable solutions are expected to have the order-parameter absolute value with only 
a single extremum inside the central lead, at the center $x=0$ between the interfaces. It should be a maximum, if 
$(df/dx)_{(l/2)-0}<0$, and a minimum otherwise. Correspondingly, the two types of symmetric solutions will be considered 
in the following. 

The solution of the first type satisfies the condition $(df/dx)_{(l/2)-0}\le 0$. It has the maximum 
$t(0)=t_d$ at $x=0$ and minima at the boundaries 
$x=\pm (l/2-0)$. The order parameter values $t_-$ 
and $t_+$ do not show up inside the central lead in this case. In accordance with the boundary conditions, 
the derivatives at the boundaries are generally nonzero and discontinuous across the interfaces. The solution of 
the second type has the minimum $t(0)=t_-$ at $x=0$ and maxima at $x=\pm (l/2-0)$, in agreement with the condition 
$(df/dx)_{(l/2)-0}>0$, while the values $f_d,\,f_+$ do not show up in the central lead.  

For the solution of the first type, one has $t_-\le t_{(l/2)-0}\le t(x)\le t_d\le t_+$ 
inside the central lead $|x|<l/2$ and gets from Eq. \eqref{RPhi89gt}:
\begin{multline}
|x|=\\ \sqrt{\dfrac{2}{t_+-t_-}}
F\left(\left.\arcsin\sqrt{\dfrac{\left(t_+-t_-\right)\left(t_d-t\right)}{\left(t_d-t_-\right)\left(t_+-t\right)}}\right|\,
\dfrac{t_d-t_-}{t_+-t_-}\right).
\label{sol1}
\end{multline}
Here the definitions of the Mathematica book are used for the notations of arguments of the elliptic integral
of the first kind $F\left(\varphi\left|\,m\right.\right)$. \cite{Wolfram2003}

Taking $x=l/2-0$ in \eqref{sol1} results in the condition associated with the central lead's length:
\begin{multline}
\!\!\sqrt{\dfrac{2}{t_+-t_-}}
F\!\left(\!\left.\arcsin\sqrt{\dfrac{(t_+-t_-)(t_d-t_{\frac{l}{2}-0})}{(t_d-t_-)(t_+-t_{\frac{l}{2}-0})}}\right|\,
\dfrac{t_d-t_-}{t_+-t_-}\!\right)\\ =\dfrac{l}{2}.
\label{sol4}
\end{multline}

The solution of the second type applies when $t_-\le t\le t_{l/2-0} \le t_d \le t_+$\, 
and \, $\bigl(g_\delta+g_\ell\bigr)f_{l/2-0}-g_\ell\cos\chi f_{l/2+0}\le 0$. For $|x|<l/2$, it takes the form 
\be
|x|=\sqrt{\dfrac{2}{t_+-t_-}}
F\left(\left.\arcsin\sqrt{\dfrac{t-t_-}{t_d-t_-}}\right|\,\dfrac{t_d-t_-}{t_+-t_-}\right).
\label{sol2}
\ee

Similarly to \eqref{sol4}, one finds from \eqref{sol2}
\begin{multline}
\sqrt{\dfrac{2}{t_+-t_-}}
F\left(\left.\arcsin\sqrt{\dfrac{t_{l/2-0}-t_-}{t_d-t_-}}\right|\,\dfrac{t_d-t_-}{t_+-t_-}\right)=\dfrac{l}{2}.
\label{sol5}
\end{multline}

The quantity $\cal E$ in the central lead can be expressed via $f_{(l/2)\pm0}$, taking $x=(l/2)-0$ 
in \eqref{RPhi88gt} and making use of (4) and (5):
\begin{multline}
{\cal E}=\biggl[1+\Bigl(g_\delta+g_\ell\Bigr)^2\biggr]f_{(l/2)-0}^2+g_\ell^2 f_{(l/2)+0}^2-\\
-2g_\ell\Bigl(g_\ell+g_\delta\Bigr)\cos\chi f_{(l/2)-0}f_{(l/2)+0}-\dfrac{1}{2}f_{(l/2)-0}^4\,.
\label{boundcond54}
\end{multline}

The order-parameter profile in the long external superconducting leads satisfies $f(-x)=f(x)$ and takes the conventional form 
(see, e.g., Refs. \onlinecite{deGennes1966,Abrikosov1988})
\begin{equation}
f(x)=f_\infty\tanh\left(\dfrac{x-(l/2)+x_0}{\sqrt{2}}\right), \quad x>\dfrac{l}{2}.
\label{external1}
\end{equation}
It has a maximum $f_\infty\le1$ at asymptotically large distances deep inside the leads and
a minimum at the pair breaking boundaries $x=\pm(l/2+0)$. The quantity $x_0>0$ is associated
with the boundary order-parameter value $f_{(l/2)+0}=f_\infty\tanh(x_0/\sqrt{2})$. The parameters $f_{(l/2)+0}$ and $f_\infty$ 
depend on the phase difference $\chi$ and the central lead's length $l$ and should be determined, together with other 
parameters of the whole solution, from the boundary and asymptotic conditions, as well as the current conservation. 

Since the derivative $df/dx$ vanishes at asymptotically large distances, it follows from (3), 
\be
i^2=(1-f_\infty^2)f_\infty^4.
\label{iext1}
\ee 
Therefore, inside the external lead, the quantity ${\cal E}_{\text{ext}}$ is conveniently associated with $f_\infty$:
\be
{\cal E}_{\text{ext}}=2f_{\infty}^2-\dfrac{3}{2}f_{\infty}^4\,.
\label{boundcond56}
\ee
One also gets from \eqref{ttt1}-\eqref{boundcond56} $t_{\text{ext},d}=t_{\text{ext},+}=f_\infty^2$ and 
$t_{\text{ext},-}=2(1-f_\infty^2)$.

Equating (5) and \eqref{iext1} results in the equation
\be
(1-f_\infty^2)f_\infty^4=g_\ell^2f_{(l/2)-0}^2f_{(l/2)+0}^2\sin^2\chi.
\label{sol3}
\ee

As the conditions $({df(x)}/{dx})\ge0$, $2\bigl(1-t_{\infty}\bigr)\le 
t_{l/2+0}\le t(x)\le t_{\infty}$ are satisfied at $x>l/2$, taking the square root
of both sides of equation \eqref{RPhi89gt} results in
\be
\dfrac{df(x)}{dx}=\dfrac{1}{\sqrt{2}f(x)}
\Bigl(f^2_{\infty}-f^2(x)\Bigr)\sqrt{f^2(x)-2\bigl(1-f^2_{\infty}\bigr)}.
\label{boundcond60}
\ee
One puts $x=l/2+0$ in \eqref{boundcond60}, substitutes (4) for the derivative
at the boundary and obtains the following relation between the parameters of the problem:
\begin{multline}
\dfrac{\left(f_{\infty}^2-f^2_{l/2+0}\right)\sqrt{2f_{\infty}^2+f^2_{l/2+0}-2}}{\sqrt{2}f_{l/2+0}}= 
\displaybreak[2] \\ =\Bigl(g_\delta+g_\ell\Bigr)f_{l/2+0}-g_\ell\cos\chi f_{l/2-0}.
\label{calem10}
\end{multline}
Positive sign of the right hand side in \eqref{calem10} agrees with the condition 
$f_{l/2-0}\le f_{l/2+0}$, which will be satisfied by the consistent values of the quantities.

Since in the absence of the supercurrent $f_\infty=1$, while for the depairing current $f_\infty^2=2/3$,
one obtains from \eqref{boundcond56}\, $\frac12\le{\cal E}_{\text{ext}}\le\frac23$. The same value of the first 
integral ${\cal E}_{\text{ext}}$ should follow from \eqref{RPhi88gt} at $x=l/2+0$. Taking into account the 
corresponding boundary condition (4) as well as (5), one gets from \eqref{RPhi88gt} in this case
\begin{multline}
\dfrac12\le\Bigl((g_\delta+g_\ell)f_{(l/2)+0}-g_\ell\cos\chi f_{(l/2)-0}\bigr)\Bigr)^2+\\ +
g_\ell^2\sin^2\chi f^2_{(l/2)-0}+f^2_{(l/2)+0}-\dfrac12 f^4_{(l/2)+0}\le\dfrac23\, .
\label{Ext2}
\end{multline}
As follows from \eqref{Ext2} and the relation $f_{l/2-0}\le f_{l/2+0}$, the boundary value $f_{(l/2)+0}=0$ is 
inappropriate for the consistent solutions discussed.

The solutions of Eq. (3), which are described by \eqref{sol1} (or \eqref{sol2}) and \eqref{external1}, and satisfy 
the boundary conditions (4), contain six parameters $t_-$, $t_d$, $t_+$, $f_{(l/2)\pm0}$ and $f_\infty$. 
The parameters are linked to each other by six equations \eqref{ttt1}, \eqref{calem10}, \eqref{sol3} and \eqref{sol4} 
(or \eqref{sol5}), where expressions \eqref{boundcond54} and (5) should be substituted for ${\cal E}$ and $i$. Joint 
solutions of the equations represent the parameters $t_-$, $t_d$, $t_+$, $f_{(l/2)\pm0}$ and $f_\infty$
as well as the whole of the inhomogeneous profile of the order parameter  \eqref{external1}, \eqref{sol1} or \eqref{sol2},
as functions of the phase difference $\chi$ and the dimensionless length of the central lead $l=L\big/\xi(T)$.
Though a numerical study of such solutions is generally required, a number of important particular problems
allow analytical descriptions. 

\section{Solutions at $\chi=0$ and $\chi=\pi$}
\label{sec: chi0pi}

When $\chi=0$ or $\pi$, the supercurrent vanishes and $t_-=0$, $t_\infty=1$, as this follows
from \eqref{sol3} and the second equation in \eqref{ttt1}. This substantially simplifies the remaining 
equations in \eqref{ttt1}, \eqref{calem10}, which allow one to express the quantities $t_d$, $t_+$ and 
$f_{(l/2)+0}$ via $f_{(l/2)-0}$:
\begin{multline}
f_{l/2+0}=\dfrac1{\sqrt{2}}\biggl[\sqrt{\Bigl(g_\delta+g_\ell\Bigr)^2+2\Bigl(1\pm\sqrt{2}g_\ell f_{l/2-0}\Bigr)}-\\ -
\Bigl(g_\delta+g_\ell\Bigr)\biggr],
\label{calem145a}
\end{multline}
\begin{multline}
t_d=1- \\ -\sqrt{\Bigl(1-f_{l/2-0}^2\Bigr)^2-2\biggl[\Bigl(g_\ell+g_\delta\Bigr)f_{l/2-0}\mp g_\ell f_{l/2+0}\biggr]^2},
\label{calem146a}
\end{multline}
\begin{multline}
t_+=1+ \\ +\sqrt{\Bigl(1-f_{l/2-0}^2\Bigr)^2-2\biggl[\Bigl(g_\ell+g_\delta\Bigr)f_{l/2-0}\mp g_\ell f_{l/2+0}\biggr]^2}.
\label{calem147a}
\end{multline}
The quantity $f_{l/2+0}$ should be considered in \eqref{calem146a} and \eqref{calem147a} as a function of $f_{l/2-0}$,
defined in \eqref{calem145a}. The upper sign in \eqref{calem145a}-\eqref{calem147a} corresponds to $\chi=0$, while the 
lower sign is associated with $\chi=\pi$. 

If $\chi=0$, it follows from \eqref{calem146a} that the equality $t_d=t_{l/2-0}$ holds, if the derivative $\frac{df}{dx}$ at 
$x=l/2-0$ in (4) vanishes, i.e., $\Bigl(g_\ell+g_\delta\Bigr)f_{l/2-0}=g_\ell f_{l/2+0}$. When the derivative is negative, one 
substitutes \eqref{calem145a}-\eqref{calem147a} and $t_-=0$ into \eqref{sol4} and obtains for the solution of the first type
the dependence $f_{l/2-0}(l)$ shown by solid curves in the left panel of Fig. 1. As seen in \eqref{sol4}, one gets 
$l\to0$ in the limit $t_{l/2-0}\to t_d$. The quantity $f_{l/2-0}(l\to0)$ describes the solid curves' starting points in the left 
panel of Fig.~1, which can be found by taking together \eqref{calem145a}-\eqref{calem147a} and~the relation 
$\Bigl(g_\ell+g_\delta\Bigr)f_{l/2-0}=g_\ell f_{l/2+0}$:
\begin{multline}
\!\!f_{\frac{l}2-0}=\dfrac{g_\ell}{\sqrt{2}\Bigl(g_\delta+g_\ell\Bigr)^2}\Biggl[\sqrt{g_\delta^2\Bigl(g_\delta+2g_\ell\Bigr)^2\!\!+
2\Bigl(g_\delta+g_\ell\Bigr)^2}-\\ -g_\delta\Bigl(g_\delta+2g_\ell\Bigr)\Biggr], \quad l\to 0.
\label{calem142}
\end{multline}
Thus, the order parameter at the boundary of the central lead remains nonzero at $\chi=0$ even in the limit $l\to0$, unless 
$g_\ell\to0$ at $g_\delta\ne0$.

A positive derivative $\frac{df}{dx}>0$ at $x=l/2-0$ takes place at a stronger suppression of the
order parameter in the central lead, that corresponds to metastable states. In the latter case one substitutes
\eqref{calem145a}-\eqref{calem147a} and $t_-=0$ into \eqref{sol5} and obtains, for the solution of the second type,
another dependence $f^m_{l/2-0}(l)$ shown by dashed curves in the left panel of Fig. (1). In accordance with
\eqref{sol5} and the condition $t_-=0$, there is the single starting point for all the dashed curves: $l\to0$ and 
$t_{l/2-0}\to0$. Furthermore, as distinct from the solid curves describing the first type of the solution, the dashed
curves take place only within a finite range of the length's values $0<l<l_{\text{max}}$, where the maximum length
at the end point is
\be
l_{\text{max}}=\dfrac{2\sqrt{2}}{\sqrt{2-f_{{l}/2-0}^2}}K\left(\dfrac{f_{{l}/2-0}^2}{2-f_{{l}/2-0}^2}\right)
\label{calem155}
\ee
and $f_{\frac{l}2-0}$ is defined in \eqref{calem142}.

As seen from \eqref{calem142} and the left panel of Fig.\,1, the quantity $f_{\frac{l}2-0}$ goes down with decreasing 
$g_\ell$, while the corresponding dashed curve adjoins the abscissa axis more and more closer. In the tunneling 
limit the latter curve fills the whole segment $0<l<l_{\text{max}}$ at $f_{\frac{l}2-0}\propto g_\ell\to0$, 
where $l_{\text{max}}\to\pi$, as it follows from \eqref{calem155}. 

Let now $\chi=\pi$. Since the relation $\frac{df}{dx}<0$ takes place in this case at $x=l/2-0$ irrespective of the relative values 
of $f_{l/2\pm0}$, one considers solely the solution of the first type and substitutes the corresponding equalities
\eqref{calem145a}-\eqref{calem147a} and $t_-=0$ into \eqref{sol4}. This results in a nonmonotonic dependence
$l(f_{l/2-0})$ and in the double-valued inverse function. The solid curves in the right panel of Fig. 1 correspond
to the energetically favorable branch $f_{l/2-0}(l)$ of the inverse function, while the dashed curves, describing a stronger 
suppression of the order parameter in the central lead, are associated with the metastable branch $f^{m}_{l/2-0}(l)$.
All the statements regarding relative energies of the states have been justified for the set of parameters studied
by numerical calculations of thermodynamic potential, which are similar to those carried out in Ref. \onlinecite{Barash2017}.

A striking difference between the curves in the left and right panels of Fig. 1 is associated, first of all, with the 
absence of any solutions in question under the condition $l<l_{\pi}$. The minimum distance $l_{\pi}$ depends on $g_\ell$ and 
$g_\delta$ and satisfies the equality $dl(f_{l/2-0})/df_{l/2-0}=0$. In the right panel of Fig. 1 $l_\pi$ represents the common 
starting point of both the solid and dashed curves at given $g_\ell$ and $g_\delta$. The metastable curves take place only 
within a finite range of the length $l_{\pi}<l<l_{\text{ps}}$. Here the maximum length $l_{\text{ps}}(g_\ell,g_\delta)$ is the 
end point of the dashed lines at given $g_\ell$ and $g_\delta$, where $f^{m}_{l/2-0}\to0$. In other words, at 
$l=l_{\text{ps}}(g_\ell,g_\delta)$ the metastable phase-slipping centers appear at the central lead's boundaries.

The quantity $l_{\text{ps}}$ is obtained after taking $f_{l/2-0}=0$ in \eqref{calem145a}-\eqref{calem147a} and substituting 
the results together with  $t_-=0$ in \eqref{sol4}:
\begin{widetext}
\be
l_{\text{ps}}(g_\ell,g_\delta)=
\dfrac{2\sqrt{2}}{\sqrt{1+\sqrt{1-g_\ell^2\Bigl[\sqrt{2+\bigl(g_\delta+g_\ell\bigr)^2}-
\bigl(g_\delta+g_\ell\bigr)\Bigr]^2}}}\,\,
K\left(\dfrac{1-\sqrt{1-g_\ell^2\Bigl[\sqrt{2+\bigl(g_\delta+g_\ell\bigr)^2}-
\bigl(g_\delta+g_\ell\bigr)\Bigr]^2}}{1+
\sqrt{1-g_\ell^2\Bigl[\sqrt{2+\bigl(g_\delta+g_\ell\bigr)^2}-\bigl(g_\delta+g_\ell\bigr)\Bigr]^2}}\right).
\label{calem102}
\ee
\end{widetext} 
In the weak-coupling limit $g_\ell\to0$ one gets $l_{\text{ps}}\to\pi$.

\section{Solutions at small distances}
\label{sec: smalll}

The solution of the problem considered can be analytically obtained at any value of $\chi$ within the zeroth-order approximation 
in a small parameter $l$, when the first argument of the elliptic integral in \eqref{sol4} or \eqref{sol5} should vanish. 
Regarding the applicability of such an approximation see the main text. For the solution of the first type one gets $t_d=t_{l/2-0}$ 
and, after substituting this in equations \eqref{ttt1}, the relation $f_{l/2-0}=\dfrac{g_\ell\cos\chi}{g_\delta+g_\ell}f_{l/2+0}$ 
follows under the condition $g_\ell\cos\chi\ge0$. Since the equality $f_{l/2+0}=0$ has been shown to be unacceptable, no solutions 
follow at $g_\ell\cos\chi<0$. For the second type's solution one obtains, in the zeroth-order in $l$, $t_-=t_{l/2-0}$, which leads 
to the same relations between $t_{l/2-0}$ and $t_{l/2+0}$.  

Substituting $\dfrac{g_\ell\cos\chi}{g_\delta+g_\ell}f_{l/2+0}$ for $f_{l/2-0}$ in \eqref{sol3} and in the boundary condition (4) 
at $x=l/2+0$, allows one to incorporate the quantities describing the central electrodes into the effective characteristics of the 
united interface with boundaries at $x=\pm l/2$ in a single symmetric Josephson junction. This results, with the phase incursion
over the central lead neglected, in (7) and in the following equality
\begin{multline}
\left(\dfrac{df}{dx}\right)_{l/2+0}\!\!\!=
\left[g_\delta+g_\ell-\dfrac{g_\ell^2\cos\chi^2}{g_\delta+g_\ell}\right]f_{l/2+0}=\displaybreak[0]
\\ =
\left[g_\delta+g_\ell-\dfrac{g_\ell^2}{g_\delta+g_\ell}+2\dfrac{g_\ell^2}{2\left(g_\delta+g_\ell\right)}
\sin^2\dfrac{\phi}{2}\right]f_{l/2+0}\, .
\label{calem92}
\end{multline}
Equation \eqref{calem92} is of the form of the boundary condition for the order-parameter absolute value 
in a single symmetric Josephson junction with the phase difference $\phi$ across the interface  
\cite{Barash2012,Barash2014_3}
\be
\left(\dfrac{df}{dx}\right)_{l/2+0}=
\Bigl(g_\delta^{\text{eff}}+2g_\ell^{\text{eff}}\sin^2\dfrac{\phi}{2}\Bigr)f_{l/2+0}.
\label{calem41}
\ee

Therefore, the problem of the double Josephson junction reduces in the limit $l\to0$ to the
behavior of a single junction. The behavior of the Josephson current flowing through a single junction 
is known in the GL theory at any coupling constants' values. Here the effective constants of the 
Josephson coupling and the interfacial pair breaking are associated with the characteristics of 
the double junction as
\be
g_\ell^{\text{eff}}=\dfrac{g_\ell^2}{2\bigl(g_\delta+g_\ell\bigr)}, \quad
g_\delta^{\text{eff}}=\dfrac{g_\delta(g_\delta+2g_{\ell})}{g_\delta+g_{\ell}}.
\label{calem42}
\ee

\providecommand{\noopsort}[1]{}\providecommand{\singleletter}[1]{#1}%

\end{document}